\documentstyle[aps,prc,epsfig]{revtex}
\newcommand{\beq}{\begin{equation}}
\newcommand{\eeq}{\end{equation}}
\newcommand{\bea}{\begin{eqnarray}}
\newcommand{\eea}{\end{eqnarray}}
\newcommand{\bef}{\begin{figure}}
\newcommand{\eef}{\end{figure}}
\newcommand{\bce}{\begin{center}}
\newcommand{\ece}{\end{center}}
\newcommand{\eg}{{\it e.g.}}
\newcommand{\ie}{{\it i.e.}}
\newcommand{\etal}{{\it et al.}}
\newcommand{\qb}{{\bar q}}
\newcommand{\xqb}{{x_{\bar q}}}
\def\lsim{\mathrel{\rlap{\lower4pt\hbox{\hskip1pt$\sim$}}
    \raise1pt\hbox{$<$}}}         
\def\gsim{\mathrel{\rlap{\lower4pt\hbox{\hskip1pt$\sim$}}
    \raise1pt\hbox{$>$}}}         
\begin{document}
%
%
\title{$D$-Meson Production from Recombination in Hadronic Collisions}
 
\author{R. Rapp$^{1}$ and E.V. Shuryak$^2$}
 
\address
{$^1$ NORDITA, Blegdamsvej 17, DK-2100 Copenhagen, Denmark\\
$^2$Department of Physics and Astronomy, State University of New York, 
    Stony Brook, NY 11794-3800, U.S.A.} 

\date{\today} 

\maketitle
 
\begin{abstract}
Nonperturbative effects in $D$-meson production in pion-nucleon and 
proton-nucleon collisions are investigated within the recombination 
model. The coalescence of perturbatively created charm quarks with  
sea- and valence-quarks from projectile and target fragments  
is shown to be competitive in magnitude with standard 
fragmentation calculations at both central (small $x_F$) and forward 
rapidities. Corresponding flavor asymmetries for inclusive 
$D$-meson production are thus mostly  
generated on the (light-) parton distribution level, and turn out to 
be in reasonable overall agreement with available fixed-target data.   
Predictions for upcoming measurements at RHIC are given.  
\end{abstract} 
\vspace{0.58cm}

\section{Introduction}

Since their discovery in the 1970's charm quarks have proved valuable 
probes of Quantum Chromodynamics (QCD) via the production of open- and 
hidden-charm states in hadronic collisions at high energy~\cite{Frixione:1997ma}. 
With the charm-quark mass, $m_c\simeq 1.5$~GeV/$c^2$, being  
on the borderline of perturbative and nonperturbative momentum
scales, mechanisms associated with both regimes are likely to play a 
role in their elementary production.
In addition, nonperturbative effects are inevitably
involved in subsequent hadronization, being related
to the wave functions of the final-state mesons or baryons. 
 
The usual way of describing charmed hadron production  
is based on the factorization theorem~\cite{qcdfac}, \ie, a perturbative 
treatment for the creation of a $c\bar c$ pair through annihilation of
partons within the incoming hadrons (folded over appropriate parton 
distribution functions), followed by individual hadronization of the 
$c$-quarks utilizing a (more or less) universal fragmentation 
function $D(z)$ (where $z$ is a normalized momentum fraction of the 
$c$-quark). Both stages in this description are not free 
of problems, or at least restricted to certain kinematic regimes. 
{\it E.g.}, in leading-order (LO) perturbative QCD (pQCD), the elementary 
$c\bar c$ production cross section underestimates open charm production 
in hadronic collisions by large factors of $\sim$5 (the so-called
$K$-factors). Next-to-leading order (NLO) estimates typically account 
for half of the deficit, with appreciable uncertainties inherent in 
underlying parameters such as  
the bare charm quark-mass, renormalization and 
factorization scales~\cite{Frixione:1997ma}. This obviously leaves 
room for  additional contributions of nonperturbative origin. 

Concerning the hadronization of $c$ and $\bar c$ quarks, some 
empirical   observations, especially at low transverse momentum 
($p_t$), cannot be captured by standard fragmentation functions. 
{\it E.g.}, flavor systematics at forward rapidities exhibit the 
so-called leading-particle 
effect~\cite{e743,na27,e653,e769-92,wa82,e791,e769-96,wa92,e791-00,selex-02}; 
that is, charmed hadrons containing light quarks 
corresponding to projectile valence quarks are much enhanced 
over their antiparticles. A natural explanation is provided by  
a {\em recombination}~\cite{DH77,DT78,TTCK79} of projectile valence 
with forward-produced (anti-) charm 
quarks~\cite{KLS81,LSS83,Hwa95,Arak98,CHM98,LS99,BJM02}, 
facilitated by small rapidity differences.       
More indirectly, it has been found~\cite{Frixione:1997ma} that 
$x_F$- and $p_t$-distributions of $D$-mesons are well reproduced by 
delta-function fragmentation functions, \ie, the produced charm quarks do 
not seem to lose momentum 
as one would expect for a (realistic) fragmentation process. 
This is suggestive for the importance of recombination mechanisms
even at central rapidities.  As opposed to high-$p_t$ emission, 
charm quarks at low $p_t$ move at low velocity relative to their
environment, with little energy available for string breaking, thus 
facilitating  a "statistical" attachment to neighboring light quarks.

Finally, as has been pointed out in 
Ref.~\cite{Frixione:1997ma}, available experimental data from both  
$p$-$N$ and $\pi$-$N$ collisions point at a significant enhancement
of $D^+/D^-$ over $D^0$ total production (dominated by central 
rapidities and low $p_t$) over expectations from perturbation theory. 
This feature is also borne out when performing empirical fits using the 
PYTHIA event generator, which require separate $K$-factors   
to fit $D^\pm$ and $D^0/\bar D^0$, differing by a factor of 
$\sim$2~\cite{pbm98}. 
 
In the present article we address the afore-mentioned flavor dependencies
by evaluating a suitably formulated recombination model for charged 
and neutral $D$-mesons extended to the central $x_F$ region. This allows 
to assess ("soft") recombination processes involving sea-quarks from both 
projectile and target nucleons (or pions), and thus aims at a consistent 
treatment of forward and total flavor production asymmetries. 
In addition, one obtains estimates for $D_s$ meson production within 
the same framework.    

The following presentation is organized as follows. 
In Sect.~\ref{sec_recom} we recall the main features of 
recombination mechanisms as have been applied in the literature
before, and generalize it for our purpose of central production. 
LO pQCD  $c\bar c$ cross sections are employed to obtain the primordial 
$c$-quark distribution, including a single (constant) $K$ factor adjusted
to the measured total yields. The two key (nonperturbative) quantities
to evaluate subsequent $D$-meson formation,  namely a two-parton 
distribution function and a recombination function,
are discussed in detail. The former provides the light quark for the  
"coalescence" process, whereas the latter characterizes the
pertinent overlap probability to form a $D$-meson. 
In Sect.~\ref{sec_data} we compare our results with available
data from both $\pi$-$N$ and $p$-$N$ reactions, and in 
Sect.~\ref{sec_rhic} we quote our predictions for upcoming 
measurements at RHIC in the $p$-$p$ mode at center-of-mass ({\it CM}) 
energy $\sqrt{s}$=200-500~GeV. 
We finish with conclusions and an outlook  for future
work in Sect.~\ref{sec_concl}. 

\section{Charm-quark recombination into $D$-mesons}
\label{sec_recom}

\subsection{General Outline of the Approach}
\label{sec_outline}
The basic idea of the recombination approach is that quarks produced
in hadronic collisions hadronize by "coalescing" with 
(anti-) quarks pre-existing in the wave function of projectile
(and/or target). This mechanism is quite different from the 
usual fragmentation, where produced quarks hadronize individually by 
string breaking, independent of their environment. Fragmentation
functions therefore ought to be universal objects, which, in principle,
can be extracted from $e^+e^-$ collisions, where 
fragmentation is expected to be the sole source of hadron production.
However, as already pointed out in the introduction, the application to
hadronic collisions is limited. 
The first clear deviations have been identified in forward production of
low-$p_t$ kaons and pions in $p$-$p$ collisions at ISR/Fermilab 
fixed target energies ($E_{lab}\ge 100$~GeV)~\cite{Ochs77}, which showed    
strong enhancement of the $\pi^+/\pi^-$ ratio at large $x_F$ 
(up to a factor $\sim$~5), and even 
more pronounced for $K^+/K^-$ . This lead Das and Hwa~\cite{DH77}
to propose, within the parton model framework, the recombination 
model: forward produced quarks preferentially "pick up" valence
up-quarks from the projectile, thus favoring the "leading" hadrons
($\pi^+=\bar d u$ and $K^+=\bar s u$) over their "non-leading" 
antiparticles.  In addition, as shown in Ref.~\cite{DT78}, 
recombination is also capable of nicely describing the observed 
increase of $\pi^-/K^-$ and $K^+/\pi^+$ for $x_F\to 1$. 

From a theoretical point of view, (anti-) charm quarks are a cleaner
probe of recombination dynamics due to a negligible probability of 
producing them in secondary reactions. Indeed, the recombination 
approach has been successfully 
applied~\cite{KLS81,LSS83,Hwa95,Arak98,CHM98,LS99,BJM02} 
(see also \cite{GV99,Carv01}) to forward production of charmed
hadrons, which similarly exhibit leading 
particle effects. However, little attention has so far been paid 
to light-flavor asymmetries in the bulk production of charmed hadrons, 
which necessarily requires substantial contributions in the 
central region ($x_F$ around 0) at low $p_t$ where most of the 
yield is concentrated.    

Here we generalize the recombination approach to incorporate  
$D$-meson formation at small $|x_F|$. 
As the low (Bjorken-) $x$ region is predominantly populated by 
sea-quarks, the main extension concerns the evaluation of their 
recombination with $c$-quarks. One of our ideas here is, that, besides 
the valence content, the proton sea possesses  a well established flavor 
asymmetry which could reflect itself in the flavor composition of 
$D$-mesons.   

To begin with, the evaluation of the production of $c\bar c$ quarks has 
to be specified. For forward $D$-meson production it has been suggested 
that an "intrinsic" charm component~\cite{Com79,Brod80} in the projectile 
wave functions could be responsible for significant recombination 
contributions~\cite{VBH92,CHM98}.
Such a "hard" charm component seemed to be required to account 
for the leading $D$-meson distributions towards large $x_F$ at ISR
energies.
On the other hand, more recent data~\cite{e769-96} indicate that the 
shape of inclusive $D$-meson $x_F$ distributions agrees well with 
pQCD predictions for bare (anti-) charm quarks (\ie, without momentum
loss due fragmentation). We take this as a motivation for   
the following picture~\cite{KLS81,LSS83} that will be employed below:  
$c\bar c$ quarks are assumed to be exclusively created in (primordial)
hard parton-parton collisions, evaluated in LO pQCD upscaled by 
a single empirical $K$ factor ($\simeq$~5-6 for typical choices of 
parton distribution functions (PDF's) and charm-quark mass $m_c=1.35$~GeV) 
to match the experimental yields. The such generated charm-quark ($x_F$-) 
distributions are then subjected to recombination 
processes which necessarily involves nonperturbative
information~\footnote{Note that in Ref.~\cite{BJM02}, a hard-scattered
(projectile-) parton itself (after radiating off a gluon that fuses into 
$c\bar c$ with a target gluon) participates in subsequent recombination, 
whereas in the present analysis spectator anti-/quarks recombine.  
Formally, the former process is suppressed by one power in $\alpha_s$, 
which, for very forward production, is compensated by a large kinematic 
enhancement. Consequently, the results of Ref.~\cite{BJM02} exhibit 
essentially no asymmetry for central $x_F\le 0.2$.}. 
For comparison with data, charm quarks that do not recombine are 
hadronized via isospin symmetric fragmentation with the usual
polarization weights (\ie, 3:1 for $D^*/D$), and, for simplicity, 
in $\delta$-function approximation for the $x_F$ dependence.  

An important part of the inclusive charged and neutral 
pseudoscalar $D$-meson yields (as reported by most experiments)  
stems from feeddown contributions from $D$-meson resonances, 
most notably the vector-mesons $D^*(2010)$. The two charge
states of the latter exhibit rather asymmetric decay characteristics: 
the $D^*(2007)^0$ decays to $\sim$~100\% in $D^0(1865)$ (with accompanying
$\pi^0$ or $\gamma$), whereas the $D^*(2010)^+$ has a branching ratio 
of 68\% into $D^0(1865)\pi^+$ and only 32\% into $D^+(1869)$ final states 
(charge conjugate modes for anti-$D$ mesons implied). 
As a consequence, the flavor composition of inclusive $D$-mesons
depends on the relative abundance of directly produced
$D$ and $D^*$ mesons (higher resonances are neglected). 
For flavor-symmetric direct production, and 
defining $\epsilon=N_{D^*}/(N_{D^*}+N_D)$, one finds for the
inclusive charged-over-neutral $D$-meson ratio
\beq
R_{c/n}\equiv \frac{(D^++D^-)}{(D^0+\bar D^0)}
=\frac{(1-\frac{2}{3}\epsilon)}{(1+\frac{2}{3}\epsilon)}  \ ,  
\eeq
whereas the inclusive particle-over-antiparticle ratios, $R_{+/-}$ and
$R_{0/\bar 0}$, are independent of $\epsilon$.   
For $c$-quark fragmentation, we employ the usual partitioning 
according to spin-isospin polarization counting implying a  
3:1 ratio for $D^*:D$, \ie, $\epsilon=3/4$ and thus $R_{c/n}=0.33$. 
On the other hand, the recombination contribution is essentially a 
coalescence process and therefore expected to be sensitive
to the mass of the product. Taking as a guideline effective thermal 
weights  as extracted from light hadron production in $pp$ 
collisions~\cite{Beca97}, with $T_H=$~170-175~MeV, results in a ratio 
$D^*/D \simeq 3 (m_{D^*}/m_D)^{3/2} \exp[-(m_{D^*}-m_D)/T_H] \simeq 3/2$, 
\ie, $\epsilon=0.6$, which will be adopted below. In the absence of 
any flavor asymmetries this value for $\epsilon$ implies $R_{c/n}=0.43$.   
The $x_F$-dependence of particle/antiparticle asymmetries is often 
displayed in terms of the so-called asymmetry function, defined via 
the production ratios as
\bea
\label{Axf}
A_{-/+}(x_F)&=&\frac{N_{D^-}-N_{D^+}}{N_{D^-}+N_{D^+}} =
   \frac{R_{-/+} -1}{R_{-/+}+1}
\\
R_{-/+}&=&\frac{1+A_{-/+}}{1-A_{-/+}} \ ,
\label{Rxf}
\eea
and likewise for $\bar D^0$ vs. $D^0$ (for baryon projectiles,
anti-$D$ mesons are the "leading" particles).

For the remainder of this section we focus on the recombination 
contribution to $D$-meson formation. In the above specified approach, 
the pertinent cross section in hadronic collisions at high energies 
takes the form~\cite{KLS81,LSS83}
\beq
x^* \frac{d\sigma^{rc}}{dx_F} = \int \frac {dx_{\bar q}}{x_{\bar q}} 
\int \frac{dz}{z} \ \left( x_{\bar q} z^* 
\frac{d^2\sigma}{dx_{\bar q} dz}\right) \ {\cal R}(x_q,z;x_F)  \ ,  
\label{dsig_rc}
\eeq
and a similar expression for $\bar D$'s upon replacing
$c\leftrightarrow \bar c$ and $\bar q \leftrightarrow q$.
Here, $x^*=E/(\sqrt{s}/2)$ and $x=p_l/(\sqrt{s}/2)$ denote the $D$-meson 
energy and longitudinal momentum, respectively, normalized to 
the incoming proton (carrying {\it CM} energy and momentum $\sqrt{s}/2$), 
$z^*$ and $z$ those of the charm-quark,
and $\xqb$ the momentum fraction of the (projectile) light antiquark
participating in the recombination (note that, for nucleons, the antiquark 
necessarily arises from the sea). The double differential cross section for
the simultaneous production of $c$ and $\qb$ quarks is given by 
\bea
x_{\bar q} z^* \frac{d^2\sigma}{d\xqb dz} =  
\int\limits_{m_c^2}^W dm_{c,\perp}^2 
\sum\limits_{i,j} \int\limits_{x_1^{min}}^{x_1^{max}} dx_1 \ 
\frac{\xqb x_1 f^{(2)}_{\qb i}(\xqb, x_1) \ x_2 f_j(x_2)}{(x_1-z_+)}   
 \ \frac{d\hat\sigma_{ij\to c\bar c}}{d\hat t}
\label{d2sig}
\eea
with kinematic boundaries $x_1^{min}=z_+/(1-z_-)$, $x_1^{max}=1-\xqb$  
(defining $z_\pm=\frac{1}{2}(z^*\pm z)$), and 
$W=s(1-z-\xqb)(1-\xqb)(1+z)/(2-\xqb)^2$.      
$d\hat\sigma_{ij}/d\hat t(\hat s, m_{c,\perp})$ denotes the 
standard LO pQCD parton fusion cross section~\cite{GOR78,Com79} 
(including the $K$-factor).
The summation over $i,j$ accounts for the possible parton
combinations from target and projectile hadrons, with $f_j$ the 
distribution function for the target (nucleon). An analogous 
expression describes recombination processes with target (light-)
quarks, upon replacing $f^{(2)}_{\qb i}(x_1) \to  f_i(x_1)$ 
and  $f_j(x_2) \to  f^{(2)}_{\qb j}(x_2)$.   
The two crucial ingredients in evaluating the cross section, 
Eq.~(\ref{dsig_rc}), are the recombination function ${\cal R}$
and the {\em two}-parton distribution function $f^{(2)}_{\qb i}$.  
In the following we discuss both entities in more detail.

\subsection{Recombination Function}
\label{sec_reco}
The recombination function ${\cal R}$ essentially represents
the (absolute value squared) wave function of the nascent charm-meson 
in terms of its quark constituents. For forward production one can 
rather straightforwardly employ the parton model in the collision 
({\it CM}) according to~\cite{DH77,Hwa95} 
\beq   
{\cal R}_D= \left[B(a,b)\right]^{-1} \ \xi_{\qb}^a \ \zeta^b 
\ \delta(1-\xi_\qb-\zeta) \     
\label{RDpm}
\eeq
with $\xi_\qb\equiv x_\qb/x$, $\zeta\equiv z/x$. 
The Beta-function prefactor, $B(a,b)\equiv \Gamma(a) \Gamma(b) 
/ \Gamma(a+b)$ ($\Gamma$: Gamma-function), ensures the correct wave 
function normalization,  
\beq 
1=\int\limits_0^1 \frac{d\xi_\qb}{\xi_\qb} \int\limits_0^1 
\frac{d\zeta}{\zeta} \ {\cal R_D} \ .  
\eeq
The actual values of the exponents $a$ and $b$ have been motivated 
in two ways.  On the one hand, in Ref.~\cite{Hwa95}, the requirements 
that the average momentum fractions of the two constituents scales
with their masses,  and that   
$G_D\equiv {\cal R}_D /(\xi_\qb \zeta)$ stays finite for $\xi_\qb\to 0$
(in the spirit of a constituent quark distribution function), 
are met by the choice $a$=1 and $b$=5. On the other hand, in 
Ref.~\cite{KLS81}, it has been argued on the basis of leading Regge 
trajectories that $a=1-\alpha_v$ with $\alpha_v$=0.5 
for $v$=$u$, $d$, and $b=1-\alpha_c$ with $\alpha_c$$\simeq$-2.2. 
Our results shown below turn out to have little sensitivity to the difference 
between these two cases.    

A more severe complication arises 
when applying the parton model wave function for small $D$-meson momenta, 
\ie, for $x_F$ close to zero (central production); here, light-cone 
momentum fractions become ill-defined, involving even contributions from 
backward moving charm or light (anti-) quarks (w.r.t. the collision {\it CM}). 
Due to the, in principle, invariant nature of the parton-model
we can resolve this problem by evaluating  the recombination function in 
a boosted frame, which we choose as the projectile rest frame. In 
practice this procedure is, however, beset with a residual frame dependence
owing to the finite transverse masses of charm- and, more sensitively, 
light-quarks. We will use the bare-charm quark mass together with 
typical light quark transverse masses of $m_{q,t}\simeq 0.3-0.5$~GeV.  
 
Another approach which avoids this complication is based on the 
formulation of the wave function in terms of rapidity variables, 
as first suggested in Ref.~\cite{TTCK79} via a (normalized) Gaussian 
distribution,
\beq
{\cal R}_D(y;y_c,y_q)=\frac{1}{\sqrt{2\pi} \sigma_y} 
\exp\left(\Delta y^2/2 \sigma_y^2\right) \ . 
\label{RDy}
\eeq
The latter is solely characterized by its width $\sigma_y$ 
and the rapidity difference $\Delta y\equiv y_c-y_\qb$ with 
the charm- and light-quark rapidities given by 
\beq
y_i=\frac{1}{2} \ln\left(\frac{E_i+p_{i,z}}{E_i-p_{i,z}} \right)
\eeq 
Again one is sensitive to the transverse mass, $m_{q,t}$ of the 
light quark within the incoming hadron, which here, however, carries  
a physical meaning: it is a nonperturbative 
quantity (indirectly) accessible by various experimental information  
(\eg, transverse momentum spectra of Drell-Yan pairs). 

In Sect.~\ref{sec_data} we will discuss both approaches of evaluating the 
recombination function, \ie, parton-model vs. rapidity-space wave functions.
The pertinent results presented turn out to be rather robust.  
This, after all, can be understood from the 
notion that also the parton model wave functions imply the 
main contribution from a rather limited relative rapidity interval
of $\langle \Delta y\rangle$=1-2~\cite{TTCK79,KLS81}.

\subsection{Two-Parton Distributions}
\label{sec_2pdf}

\subsubsection{Kinematic Correlations}
\label{sec_corrkin}
The two-parton distribution function (2-PDF) $f^{(2)}_{\qb i}$, which 
importantly figures into the double-differential cross section, 
Eq.~(\ref{d2sig}),  characterizes the probability to simultaneously 
find parton $i$ at $x_1$ (participating in the hard fusion process
into $c\bar c$) and
the antiquark $\bar q$ with momentum fraction $\xqb$ (participating
in the recombination process). In general it can be cast into the
form
\beq
\xqb x_1 f^{(2)}_{\qb i}(\xqb, x_1)= \xqb \hat{f}_{\qb}(\xqb) \ x_1 f_i(x_1) 
 \  \rho(\xqb,x_1) \ , 
\eeq 
where $\hat{f}_{\qb}(\xqb)$ represents the constrained probability
for parton $\qb$ at momentum $\xqb$ under the condition that parton $i$ is 
at $x_1$, and $\rho(\xqb,x_1)$ encodes all other correlations which 
are usually assumed to be governed by phase space.  

As a simple ansatz the following factorized form has been 
proposed~\cite{DH77}, 
\beq
f^{(2)}_{\qb i}(\xqb, x_1)= C_2 \ f_\qb(\xqb) \ f_i(x_1) \ (1-\xqb-x_1)^p \ , 
\label{facto_0}
\eeq 
where $C_2$ is a normalization constant. The last factor in 
Eq.~(\ref{facto_0}) implements a phase space dependence for large 
momentum fractions $\xqb+x_1\to 1$ (for a collinear 3-body phase 
where, in the {\it CM} system, target and forward $[i\qb]$ state recede 
from each other with the projectile at rest, one has $p=1$~\cite{DH77}).  
Some confidence in this ansatz was drawn 
from phenomenological successes in describing, \eg, (semi-) inclusive 
$\pi$- and $K$-production in the fragmentation region~\cite{DH77,DT78}.
In subsequent work~\cite{Hwa95} the 2-PDF has been elaborated on a 
more microscopic basis within the so-called valon model of (low-$p_t$) 
hadronic structure, where light (anti-) quark distributions 
are generated from composite valence quarks at low momentum scales, see, \eg, 
Ref.~\cite{HY02a} for a recent update.
Within this approach, also $c\bar c$ production arises from  
Fock components of the valon distributions, similar 
to the "intrinsic-charm" model~\cite{Com79,Brod80}. Their contribution 
can be significant at large $x$, but cannot account
for inclusive charm yields. 
However, since our primary objective here is the evaluation of $D$-meson 
flavor dependencies in {\em bulk} production, we retain the $c\bar c$ cross 
section as a hard process in factorized form, cf.~Eq.~(\ref{d2sig}). Thus,   
results to be discussed below will refer to the simplified ansatz, 
Eq.~(\ref{facto_0}), for the 2-PDF's.  
  
A more explicit treatment anti-charm quark recombination with leading 
(valence) light quarks within a factorized scheme has 
been undertaken in Refs.~\cite{KLS81,LSS83}. At the price of more 
schematic underlying 1-PDF's, a 2-PDF of the form 
\beq
f^{(2)}_{vi}  =A \ x_v^{-\alpha_v} \ x_1^{-\gamma} \ (1-x_1-x_v)^m  
         \ (1-x_1)^k   
\label{f_kls}
\eeq
has been proposed. Here,  $x_v$ denotes the momentum fraction of 
the (recombining) valence quark in the nucleon, and the exponents 
$\alpha_v$, $\gamma$, $k$ and $m$ reflect the appropriate limiting 
behaviors for $x_{1,v}\to 0,1$ (\eg, $\alpha_v=0.5$ for valence
quarks, $\gamma=1$ for sea quarks and gluons).  
The ansatz, Eq.~(\ref{f_kls}), is furthermore constrained by the 
normalization condition 
\beq
\int\limits_0^{(1-x_1)} dx_v \ f^{(2)}_{vi}(x_1,x_v) = f_i(x_1) \ ,  
\label{norm}
\eeq
which can be used to fix the numerical constant $A$  as well as
$k$ and $m$ according to the parton type $i$ (note that this can only 
be applied for the case of valence quarks $v$). 
In particular, it has been argued~\cite{KLS81} that for fixed 
$x_v$, $f^{(2)}_{vi}(\eta_1)$ with $\eta_1\equiv x_1/(1-x_v)$ can be 
interpreted as the single-parton distribution of the quark $i$ 
in the hadron remnant $\tilde{h}_1$ (with the
valence quark $v$ removed), and that for $x\to 1$ and $z\to x$ an 
approximate factorization into two single-parton distributions holds,
\ie,
\bea
f^{(2)}_{vi} & \simeq & C \ x_v^{-\alpha_v} \ (1-x_v)^{m-\gamma} \ f_i(\eta_1) 
\nonumber\\
           & \propto &  f_v(x_v) \ (1-x_v)^{-\gamma}  \ f_i(\eta_1)  \  .
\eea

In an alternative view, one might consider $f^{(2)}_{vi}$ at fixed 
$x_1$ (being determined on a shorter timescale --  hard $c\bar c$ production 
-- than $x_v$) as the single-parton
distribution of quark $v$ in the hadron remnant without parton $i$.
In this case, one straightforwardly obtains 
from Eq.~(\ref{f_kls}) the factorized form
\beq
f^{(2)}_{vi}=\tilde{A} \ f_v(\eta_v) \ f_i(x_1) \ (1-x_1)^{-1} 
\label{facto_v}
\eeq
with $\eta_v\equiv x_v/(1-x_1)$ and $\tilde{A}=1$ upon imposing 
Eq.~(\ref{norm}). This result suggests an immediate 
generalization for sea quarks according to
\beq
f^{(2)}_{\qb i}=f_\qb(\eta_\qb) \ f_i(x_1) \ (1-x_1)^{-1} \  
\label{facto_s}
\eeq  
with $\eta_\qb\equiv \xqb/(1-x_1)$.  
    
To illustrate the differences between the simple factorization ansatz 
with phase space correction, Eq.~(\ref{facto_0}), and the newly 
suggested ansatz "remnant" 2-PDF, 
Eq.~(\ref{facto_s}), we show in Fig.~\ref{fig_App}  partice-antiparticle
asymmetries in $D$-meson production for fixed target $p$-$p$ collisions. 
The results include the recombination contribution only, and are based 
on a rapidity-space recombination function, Eq.(\ref{RDy}), 
with $\sigma_y=0.5$ (using GRV94-LO~\cite{GRV95} for the underlying
1-PDF's).   In the central region, where both $\xqb$ and $x_1$ are 
small, mutual (anti-) correlations are suppressed, and the two 
ans\"atze indeed approximately coincide.  
However, the  "remnant"-2-PDF, Eq.~(\ref{facto_s}), develops 
a somewhat steeper increase of the leading-particle asymmetry 
with increasing $x_F$. Note also, that recombination with 
valence quarks is the prevalent process even for central $x_F$
(indicated by nonzero values at $x_F=0$, as well as the 
larger asymmetry in the neutral ratio involving $u$-quarks).  
The $\bar d /\bar u$ asymmetry in the proton sea, which 
favors $D^+$ formation, does not significantly suppress the 
charged ratio. One should also point out that the 
absolute magnitude of the recombination cross section
for the individual $D$-meson states is about 20\% larger
with the 2-PDF of Eq.~(\ref{facto_s}) (which leads to larger
total asymmetries once combined with isopsin-symmetric
fragmentation contributions).  
\bef[!h]
\vspace{-0.8cm}
\bce
\epsfig{file=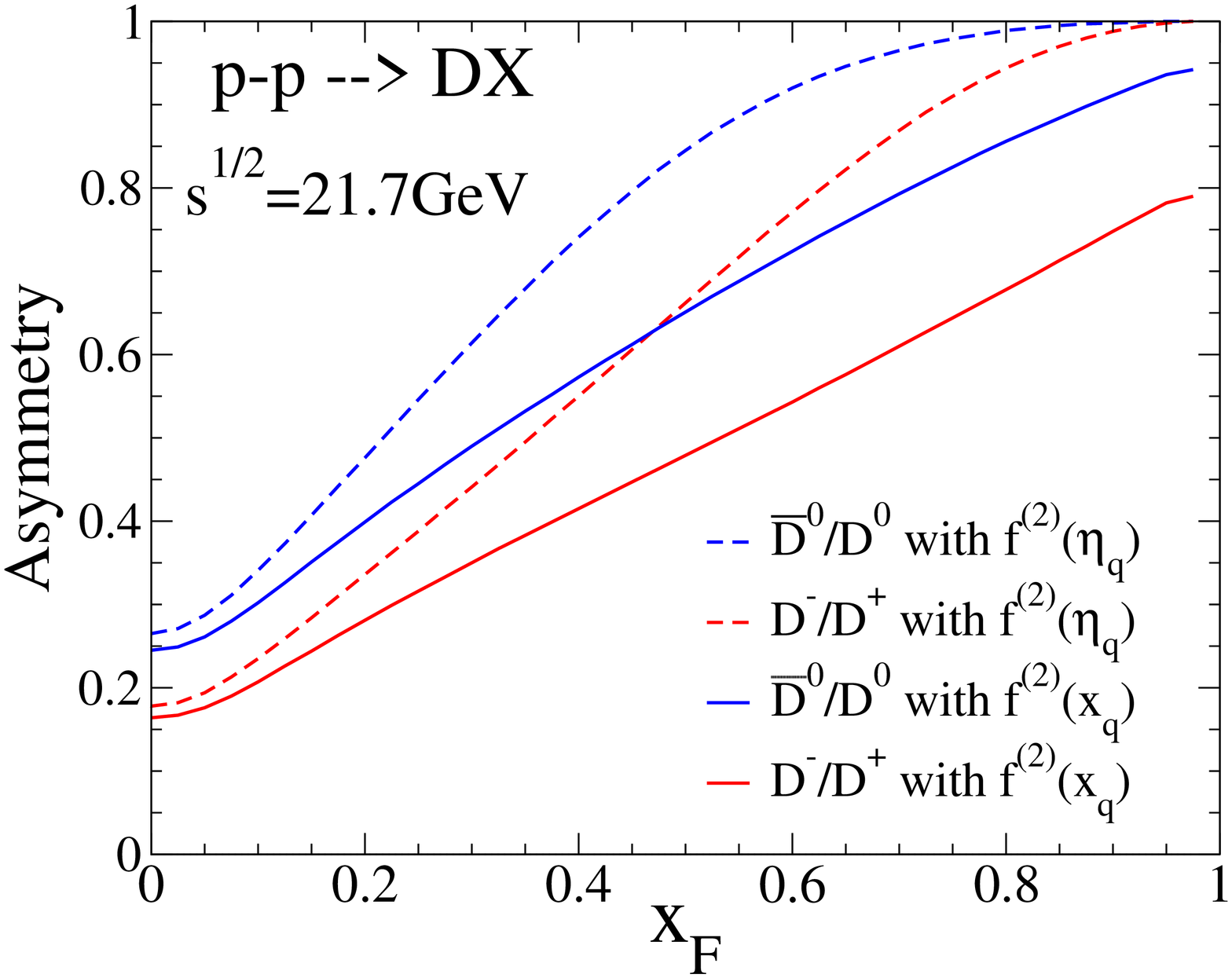,width=8cm,angle=-90}
\ece
\caption{Asymmetry function $A(x_F)$, Eq.~(\protect\ref{Axf}),
for $\bar D$/$D$ production from the recombination part of the 
cross section only in  $p$-$p$ collisions at $E_{lab}=250$~GeV. 
The dashed lines are obtained by using "remnant"-2-PDF's, 
Eq.~(\protect\ref{facto_s}) for both valence and sea-quarks,
whereas the full lines correspond to the factorized 2-PDF, 
Eq.~(\protect\ref{facto_0}) with $C_2=p=1$.
The respective upper (lower) lines are for the $\bar D^0 /D^0$ 
($D^-/D^+$) asymmetries.}
\label{fig_App}
\eef

\subsubsection{Flavor Correlations}
\label{sec_corrflav}
An obvious (global) flavor correlation that will be 
imposed throughout consists of "vetoing" a light (anti-) quark 
flavor $i$ for the recombination process if this type has 
been used in the hard process of creating a $c\bar c$ pair,    
\ie, $\bar q \ne i$. The significance of this anti-correlation
mainly establishes itself for leading $D$-mesons ($c\bar q$) 
in forward production with meson beams~\cite{Tash02}, where the 
valence anti-quark provides a large part of the $c\bar c$ yield.  

A more delicate issue is the one of local correlations, 
relating to anisotropies of the quark and gluon distributions 
within a hadron. To illustrate their possible effects we will
investigate one rather extreme scenario, which can be motivated, 
\eg, by the importance of instantons in hadron structure.
The latter provide a qualitative explanation for the observed 
$\bar d /\bar u$ excess in the proton~\cite{Fort89,DK91}
(cf.~Ref.~\cite{Peng01} for a recent overview), as follows. 
Due to the specific properties of the instanton-vertex, a valence 
$u$-quark coupling to it has to be accompanied by an in- 
and outgoing $d$-quark. The $d$-quark line can either be provided 
by another valence-quark (which generates a strong $ud$-diquark 
binding in the nucleon), or be closed off by a condensate
insertion (which, after all, generates the constituent mass
of the $u$-quark). The second case implies that  $u$-valence quarks 
are essentially accompanied by a $d\bar d$ "cloud", and $d$-valence
quarks by a $u\bar u$ one, 
thus generating a 2:1 asymmetry in $\bar d /\bar u$, 
which roughly corresponds to the experimental value 
in the $x\le 10^{-1}$ region. At the same time, one 
expects the gluon cloud to be concentrated around 
the valence quarks. Imposing such a flavor 
correlation on the 2-PDF (\eg, for gluon fusion, 
in 2/3 of the cases $c$ and $\bar c$ can only recombine 
with either $u$ valence or $d$ and $\bar d$ sea-quarks),
will be referred to as "local flavor correlations" 
below.

\section{Comparison with Experiment}
\label{sec_data}

\subsection{Total $c\bar c$ Yields and $K$-Factors}
\label{sec_Kfac}
It is well-known that LO pQCD calculations for $\bar q q , gg \to c\bar c$,
coupled with LO PDF's for the colliding hadrons, underestimate total 
charm production by an appreciable magnitude characterized by a typical
factor $K\simeq$~5. The latter depends rather little on the specific parton 
distribution function, but bears some sensitivity to the value 
of the charm-quark mass. NLO results can in principle account 
for a substantial part of the discrepancy, but large uncertainties 
due to the choice of renormalization and factorization scales 
persist~\cite{Frixione:1997ma} (see also Ref.~\cite{Vogt02} for a 
recent update). We here adopt the LO results for 
$d\hat\sigma_{ij\to c\bar c}/d\hat t$ in 
Eq.~(\ref{d2sig}), with the (bare) charm-quark mass fixed at 
$m_c=1.35$~GeV.  For comparison we will allow for two different sets of 
PDF's in our calculations, \ie, the GRV-94 LO~\cite{GRV95} and 
MRST01~\cite{MRST01}\footnote{the 
MRST01 parametrization is performed to higher order (HO) in $\alpha_s$, 
which is, strictly speaking, not consistent with the LO evaluation of 
the $c\bar c$ cross section; on the other
hand, nonperturbative effects introduced via the recombination 
mechanism evade any expansion in $\alpha_s$ for the $D$-meson 
production cross section of Eq.~(\ref{d2sig}).} sets for nucleons, 
as well as GRS-99 LO~\cite{GRS99} for pions. 
Fig.~\ref{fig_sigccb} shows the fit to fixed target $pN$ (left panel) 
and $\pi N$ (right panel) reactions, which are reasonably well 
reproduced, albeit with significantly different $K$-factors (which 
might be related to the fact that for pion projectiles 
valence-antiquarks contribute appreciably through
annihilation on target valence-quarks).
\bef[!h]
\vspace{-1.0cm}
\bce
\epsfig{file=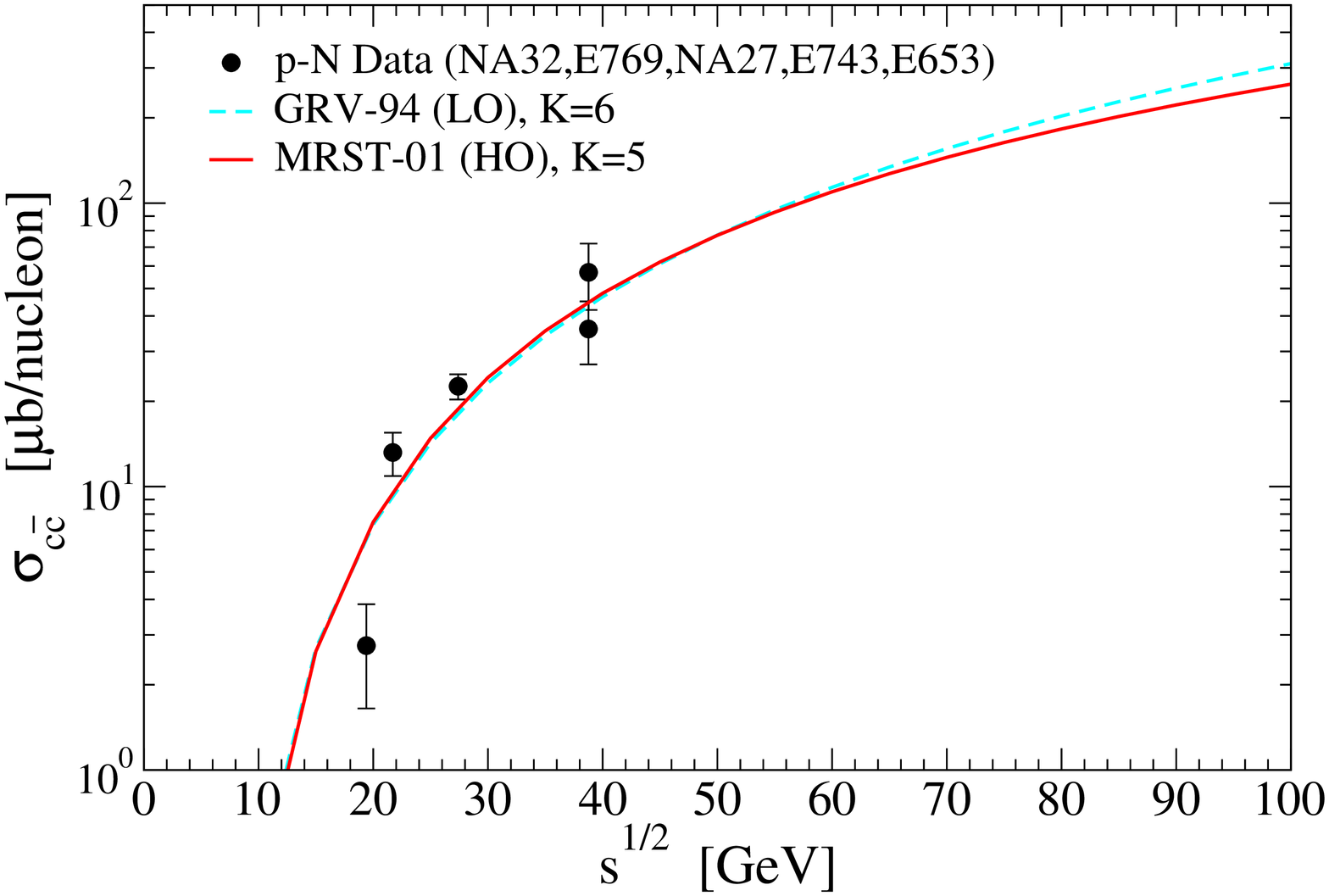,width=6.85cm,angle=-90}
\epsfig{file=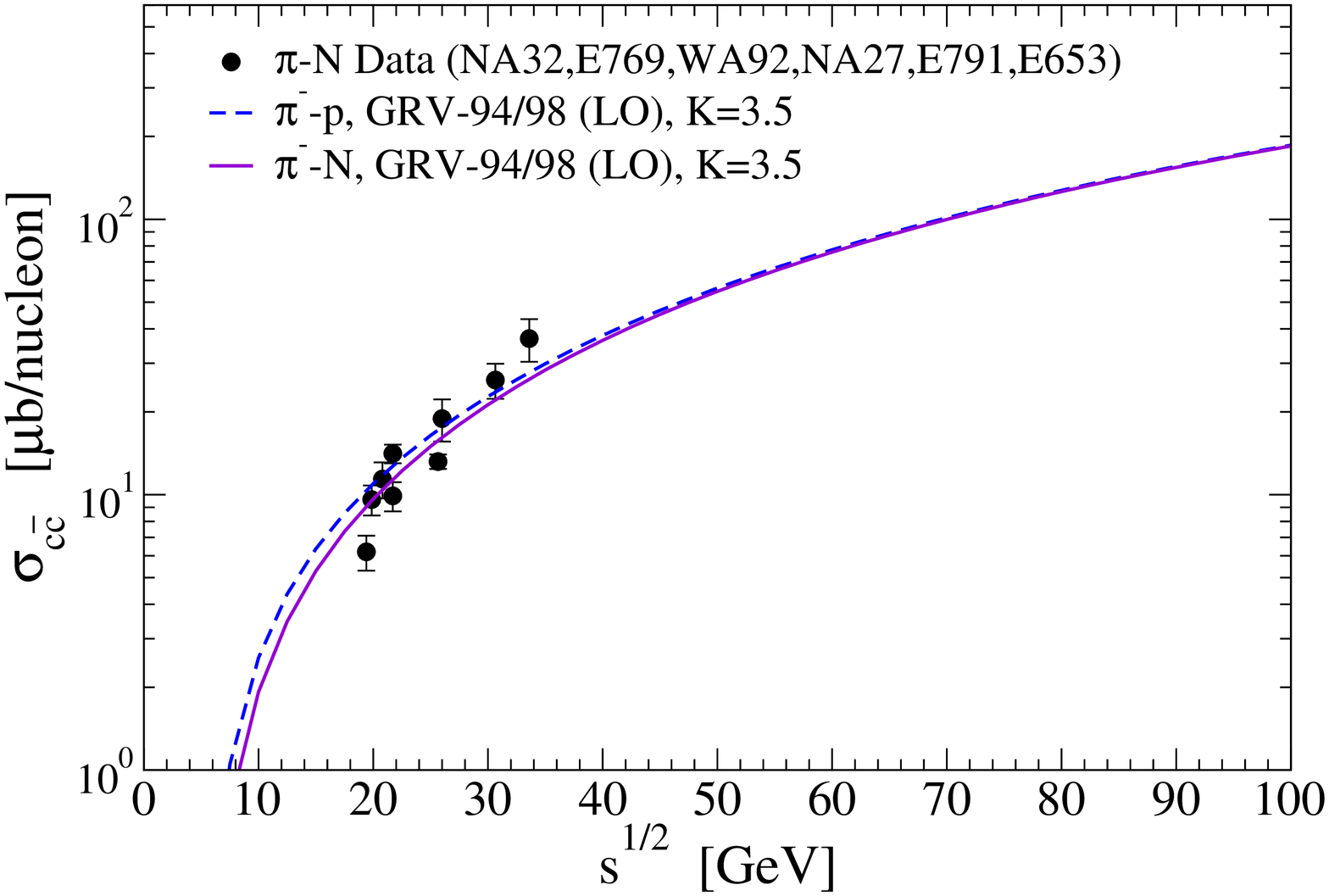,width=6.85cm,angle=-90}
\ece
\caption{Total $c\bar c$ production cross 
sections~\protect\cite{na32-88,e743,na27,na32-91,e653,e769-96} in $pN$
(left panel) and $\pi N$ (right panel) collisions compared to LO pQCD 
calculations upscaled by the indicated $K$-factors. 
Note that following Ref.~\protect\cite{wa75-92}, the experimental 
$D+\bar D$ cross sections have been multiplied by a factor of 
1.5 to account for $D_s$ mesons (20~\% of $D$) and charmed baryons (30~\% 
of $D$, most notably $\Lambda_c$).}
\label{fig_sigccb}
\eef

Based on the these total $c\bar c$ cross sections we will in 
the following address the magnitude and flavor composition of $D$-meson
production from recombination based on Eqs.~(\ref{dsig_rc}) and (\ref{d2sig}). 
As detailed in Sect.~\ref{sec_recom}, charm quarks that do not recombine 
will be hadronized with "standard" fragmentation assuming no longitudinal
momentum loss. We will mainly focus on $x_F$ distributions and 
excitation functions (\ie, inclusive yields for $x_F>0$ as a function
of collision energy, $\sqrt s$) for the ratios $(D^+ +D^-)/(D^0+\bar D^0)$, 
$D^-/D^+$ and $\bar D^0 /D^0$.

\subsection{Flavor Asymmetries I: Proton-Nucleon Collisions}
\label{sec_pN}

\subsubsection{Inclusive $x_F$ Distributions}
Let us first address the typical features of recombination cross sections 
as a function of $x_F$. Fig.~\ref{fig_e769} confronts E769 data~\cite{e769-96} 
for inclusive $D$+$D_s$ mesons with our calculations for recombination
with different wave functions ${\cal R}_D$ according to Eqs.~(\ref{RDpm}),
(\ref{RDy}), and a common two-parton distribution function  
of factorized form, Eq.(\ref{facto_0}), with $C_2^N=p_N=1$ (at this
point there is no sensitivity to more detailed properties of the 2-PDF).
Also shown is (80\% of) the pQCD yield for $c$ and $\bar c$ quarks 
(solid curve), assumed to convert into $D$-mesons without longitudinal 
momentum loss ($\delta(x-z)$-fragmentation). This distribution (for the MRST01
set with $K$=5 as fixed above) reproduces the data reasonably well.
In general, the fraction due to recombination is sizable in the central 
region (where it makes up  30-60\% of the total cross section), and
starts to dominate at forward  $x_F\ge 0.4$, where it can even exceed
the primordial $c$-quark distribution by "pick-up" of a comoving
light quark.  

In more detail, we find that using the parton-model recombination function
with $a$=1, $b$=5 and $m_t=0.3$~GeV gives a $\sim$50\% contribution to the
total $D$-meson yield. This is reduced to $\sim$30\% upon using
$m_t=0.5$~GeV (not shown), with most of the reduction occurring at central
$x_F\le 0.2$.  On the contrary, with the rapidity space recombination 
function the sensitivity to the transverse mass is essentially absent, 
giving identical yields within 2\% for the two values 
\bef[!ht]
\vspace{-0.8cm}
\bce
\epsfig{file=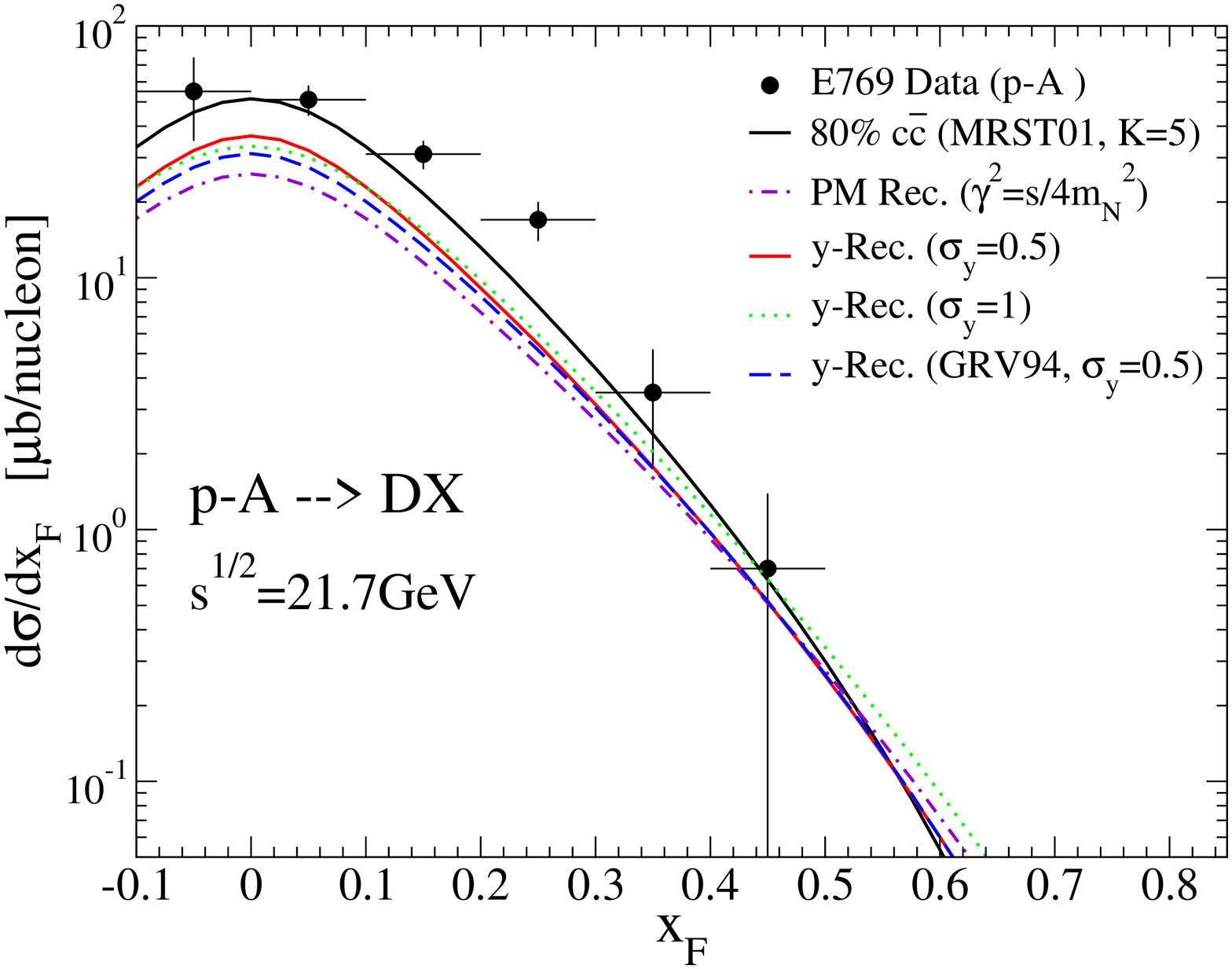,width=8cm,angle=-90}
\ece
\caption{Comparison of E769 $p$-$A$ data for inclusive production of
$D$ and $D_s$ mesons to the calculated recombination contributions
using the parton-model wave function (dashed-dotted line), 
Eq.~(\ref{RDpm}) with $a$=1 and $b$=5, as well as the rapidity-space ones, 
Eq.~(\ref{RDy}) with $\sigma_y$=0.5 (solid line) and 1 (dotted line).  
The long-dashed line is obtained from the solid one upon replacing 
the MRST01 PDF's by the GRV94-LO set.}
\label{fig_e769}
\eef
\noindent
of $m_t$.  When increasing $\sigma_y$ from, 
\eg, 0.5 to 1, the $x_F$ distribution broadens slightly with the integrated
yield again being stable (within 1\%). Also, the use of rapidity wave functions 
increases the recombination fraction of $c$ and $\bar c$ quarks to almost 70\%.  
Due to these rather robust features of ${\cal R}_D(\Delta y)$,
(and the absence of boost ambiguities for central $x_F$ as opposed
to the parton model wave function , cf.~Sect.~\ref{sec_reco}) 
we will employ this form from now on (fixing $\sigma_y=0.5$) unless 
otherwise stated. 

\bef[!ht]
\vspace{-0.8cm}
\bce
\epsfig{file=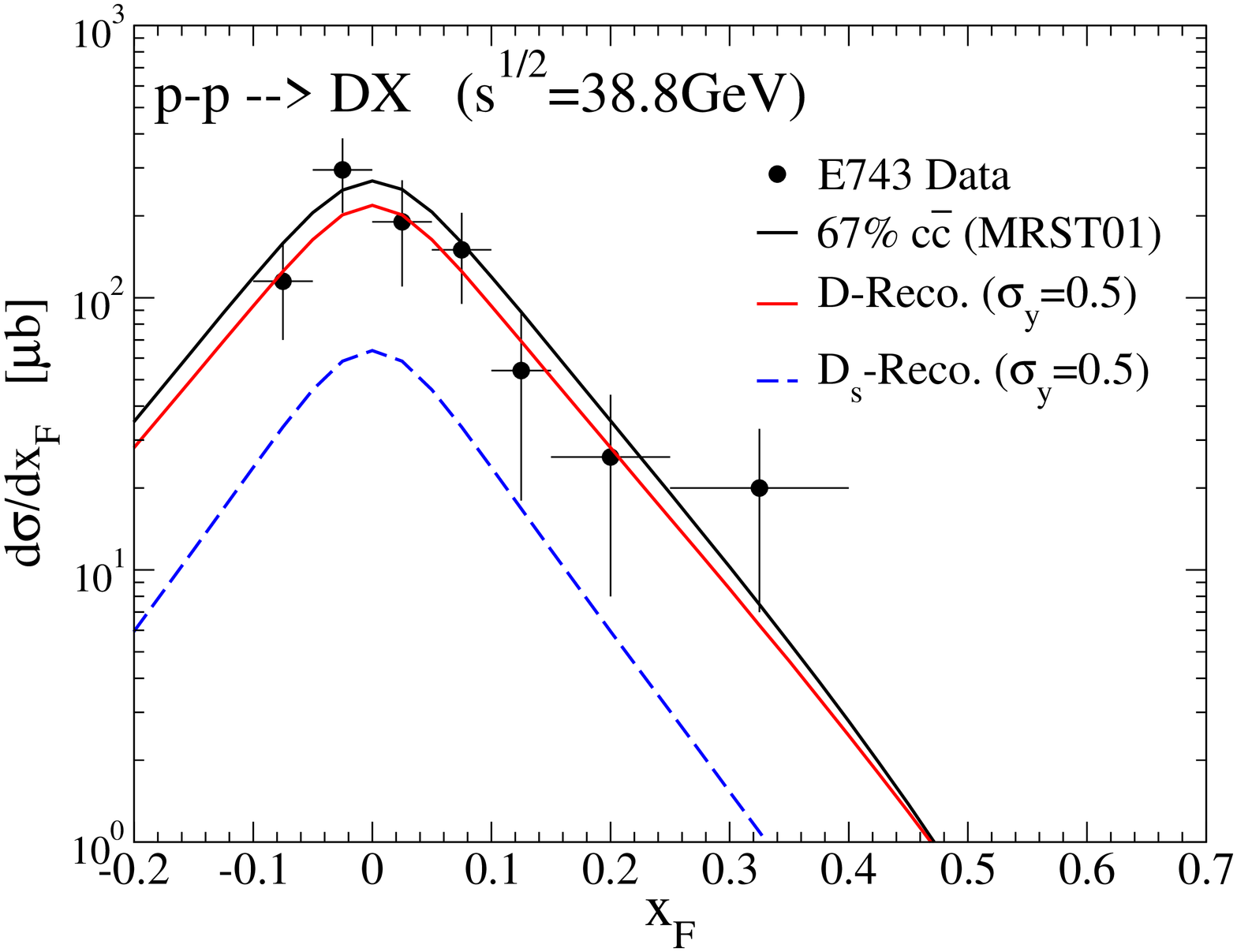,width=8cm,angle=-90}
\ece
\caption{Comparison of E743 $p$-$p$ data for inclusive production of
nonstrange $D$ mesons to the recombination contributions calculated with 
the rapidity wave function, Eq.~(\ref{RDy}), with $\sigma_y$=0.5 
(lower solid line). The upper solid line corresponds to 67\% of the underlying 
$c$ and $\bar c$ quark distributions (MRST01 with $K$=5).   
The dashed line indicates the recombination cross section into $D_s$-mesons.}
\label{fig_e743}
\eef
A similar comparison at  higher energy is performed in Fig.~\ref{fig_e743}
with nonstrange $D$-mesons from $p$-$p$ collisions measured by E743~\cite{e743}.
At $\sqrt{s}$=39~GeV, the fraction of the recombination cross section
to the expected integrated yield (which corresponds to 2/3 of the total
$c\bar c$ cross section) amounts to almost 80\%, as compared to about
70~\% for the corresponding calculation at $\sqrt{s}$=21.7~GeV,
cf.~Fig.~\ref{fig_e769}. This is due to the fact that at higher energies
smaller $x$ values of the parton distributions are probed where
the occupancy is higher. Of course, for recombination fractions
approaching 1 interference effects will play an increasingly important
role, which have been neglected here. We also note that the recombination
cross section into strange $D_s$-mesons, displayed by the dashed line
in Fig.~\ref{fig_e743}, constitutes about 25\% of the nonstrange one, 
very similar to what has been inferred for total yields~\cite{wa75-92}. 
This fraction is somewhat smaller when using the GRV94-LO distributions. 

\subsubsection{Excitation Function of Flavor Ratios}
From here on all our results for $D$-meson recombination cross sections are 
supplemented with (isospin-symmetric) $\delta$-function fragmentation
of the remaining $c$ and $\bar c$ quarks (except at large $x_F$ where
the recombination part may even exceed the $c$-quark distribution), 
with appropriate feeddown systematics from $D^*$ resonances, 
as elaborated in Sect.~\ref{sec_outline}. Throughout, the two parameters
of the nucleon 2-PDF, Eq.~(\ref{facto_0}), are fixed at $C_2$=0.8 and $p$=1, 
together with rapidity space recombination functions ($\sigma_y=0.5$).
For the underlying 1-parton distribution functions we choose the 
GRV94-LO set (in order to be consistent with 
the GRS99~\cite{GRS99} pion sets in subsequent sections).   
For nuclear targets, the appropriate neutron fraction has been employed.  
\bef[!b]
\vspace{-0.5cm}
\bce
\epsfig{file=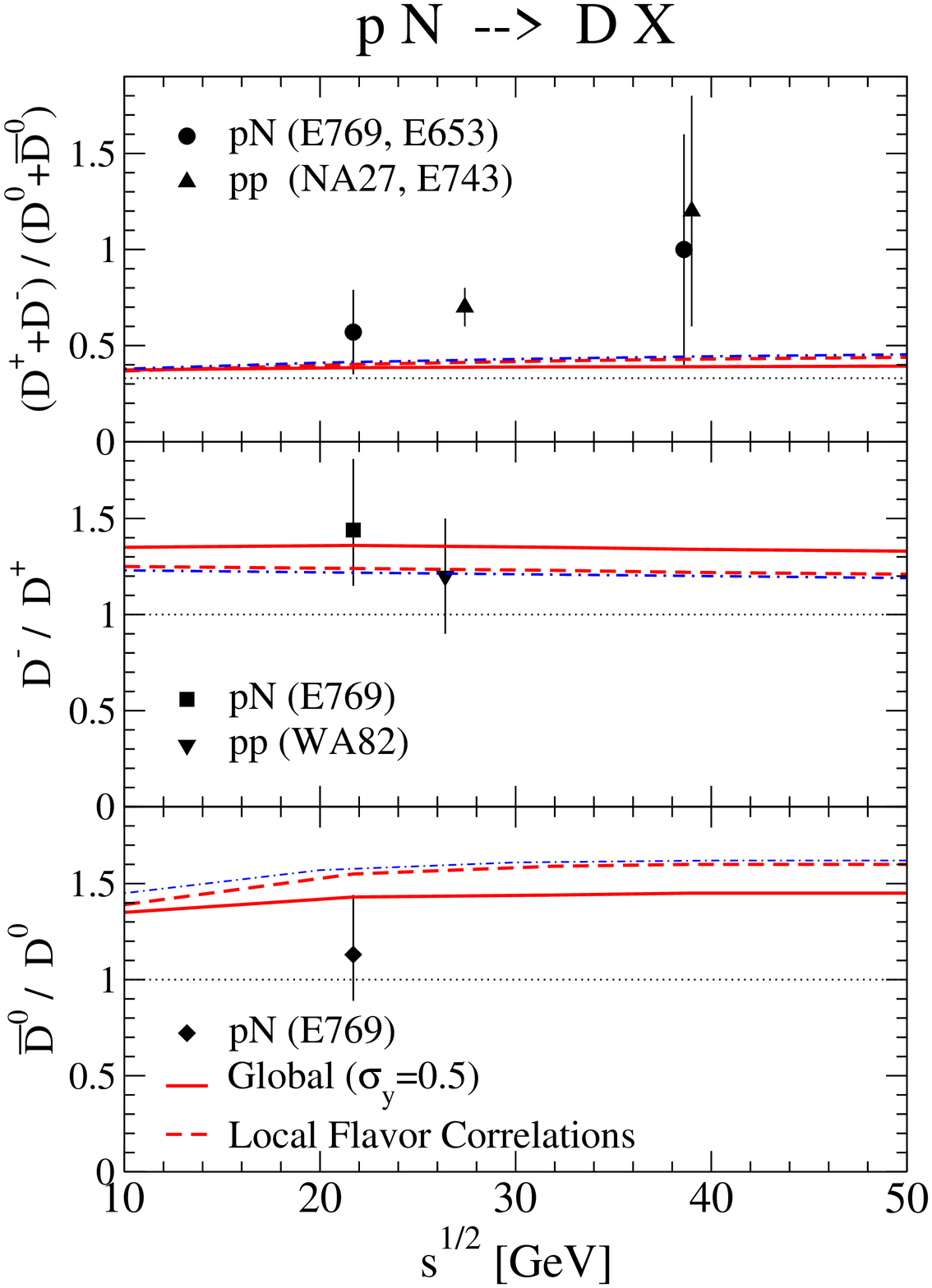,width=9cm,angle=0}
\ece
\caption{Collision energy dependence of various $D$-meson flavor
ratios in $pp$ and $p A$ collisions compared to model calculations
accounting for both recombination and isospin-symmetric 
fragmentation contributions (including feeddown from $D^*$ mesons using 
recent banching ratios). Rapidity space recombination functions (with
$\sigma_y$=0.5) have been employed in connection with the factorized 2-PDF,
Eq.~(\protect\ref{facto_0}), with $C_2$=0.8 and $p$=1. The results for 
$pN$ collisions ($Z/N$=0.8 for target nucleons) are given by 
the solid and dashed lines, where the latter include an additional 
assumption of local flavor correlations, 
cf.~Sect.~\protect\ref{sec_corrflav}. 
The dashed-dotted lines are for $pp$ collisions including
local flavor correlations as well.  
The dotted lines represent the isospin-symmetric fragmentation mechanism 
alone (assuming a $D^*/D$ ratio of 3).}
\label{fig_ratpN}
\eef
Fig.~\ref{fig_ratpN} shows the pertinent $\sqrt{s}$-dependence of
$x_F$-integrated nonstrange $D$-meson flavor ratios in $pp$ and $pN$
collisions, indicating reasonable agreement with 
available measurements for charged and neutral ratios (middle and lower
panel, respectively), whereas the asymmetry in the charged
over neutral ratio (upper panel) appears to be underpredicted.
A marginal increase of the latter can be achieved by introducing local
flavor correlations within a valence-quark type picture (as outlined
in Sect.~\ref{sec_corrflav}), but not enough to significantly improve
the description of the data. For pure proton targets, an additional small
increase for the charged over neutral ratio is found (dashed-dotted
lines). In fact, for proton targets the increase in the individual
recombination with both valence $u$-quarks to form $\bar D^0$ mesons
and with sea $\bar d$-quarks to form $D^+$ mesons is very moderate, as
indicated by the $D^-/D^+$ and $\bar D^0/D^0$ ratios on proton targets
(cf.~dashed dotted lines in the middle and lower panels).

\subsection{Flavor Asymmetries II: Pion-Nucleon Collisions}

\subsubsection{$x_F$ Dependencies}
Flavor asymmetries in forward $D$-meson production from $\pi N$ collisions
are experimentally well-established~\cite{e769-92,wa82,e791,wa92}.
In Fig.~\ref{fig_Awa92} we compare results of our approach with
the $x_F$-dependence of $D^-/D^+$ and $D^0/\bar D^0$ ratios as measured 
by WA92~\cite{wa92} in 350~GeV $\pi^-N$ reactions. 
\bef
\vspace{-0.4cm}
\bce
\epsfig{file=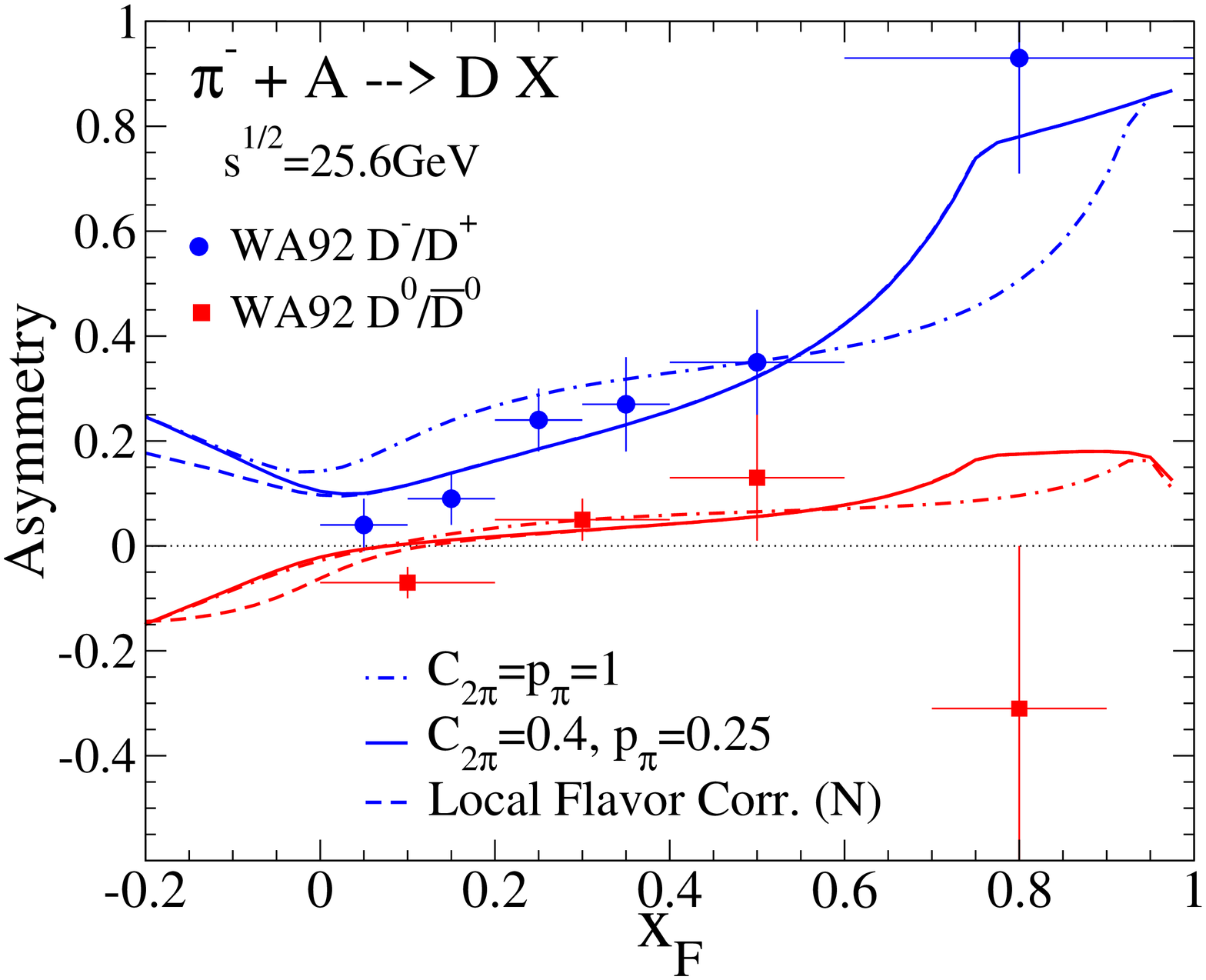,width=8cm,angle=-90}
\ece
\vspace{-0.2cm}
\caption{WA92 data~\protect\cite{wa92} for the inclusive $D^-/D^+$ 
(circles) and $D^0/\bar D^0$ (squares) asymmetries versus Feynman-$x_F$ in 
350~GeV $\pi^- A$ collisions, compared to recombination model calculations 
supplemented with isospin-symmetric $\delta$-function fragmentation 
contributions.}
\label{fig_Awa92}
\eef
The simple factorization for the 2-PDF, \ie, Eq.~(\ref{facto_0})
with $C_2=p=1$ for pions, leads to a $x_F$-dependence of the charged 
ratio (upper dashed-dotted line) that is somewhat flatter than the data. 
The description is much improved (upper solid curve) when employing a 
weaker phase space suppression (reducing the exponent to, \eg, $p_\pi$=1/4) 
in connection with a smaller normalization constant (\eg, $C_2^\pi=0.4$). 
The latter suppresses the asymmetry in the central region, whereas the 
former entails a stronger increase towards forward $x_F$.     
On the other hand, the neutral $D$-meson ratio is insensitive 
to this modification. The main feature of these data is the  
near absence of a leading particle asymmetry, even at very forward 
$x_F$. The model calculations essentially reproduce this
behavior, which arises from the fact that the $\bar u$-valence
quark in the pion, which on average carries a large projectile
momentum fraction, contributes significantly to the (hard)
$c\bar c$ production process (via annihilation on $u$-valence
quarks in the nucleon), and thus is not at disposal for subsequent
recombination.  For small $x_F$ the measured asymmetry seems to 
become even negative, which in the calculations is due to significant 
contributions from recombination processes of $\bar c$ quarks with $u$ 
quarks from target nucleons. This effect becomes more pronounced when
introducing local flavor correlations within the nucleon (lower dashed 
line), which increases (decreases) recombination with $u$ ($\bar u$) 
into $\bar D^0$ ($D^0$) mesons; at the same time, the charged ratio 
becomes somewhat suppressed at negative $x_F$ (upper dashed line), 
due to reduced (enhanced) recombination with $d$ ($\bar d$) quarks
into $D^-$ ($D^+$).   

\bef[!t]
\vspace{-0.4cm}
\bce
\epsfig{file=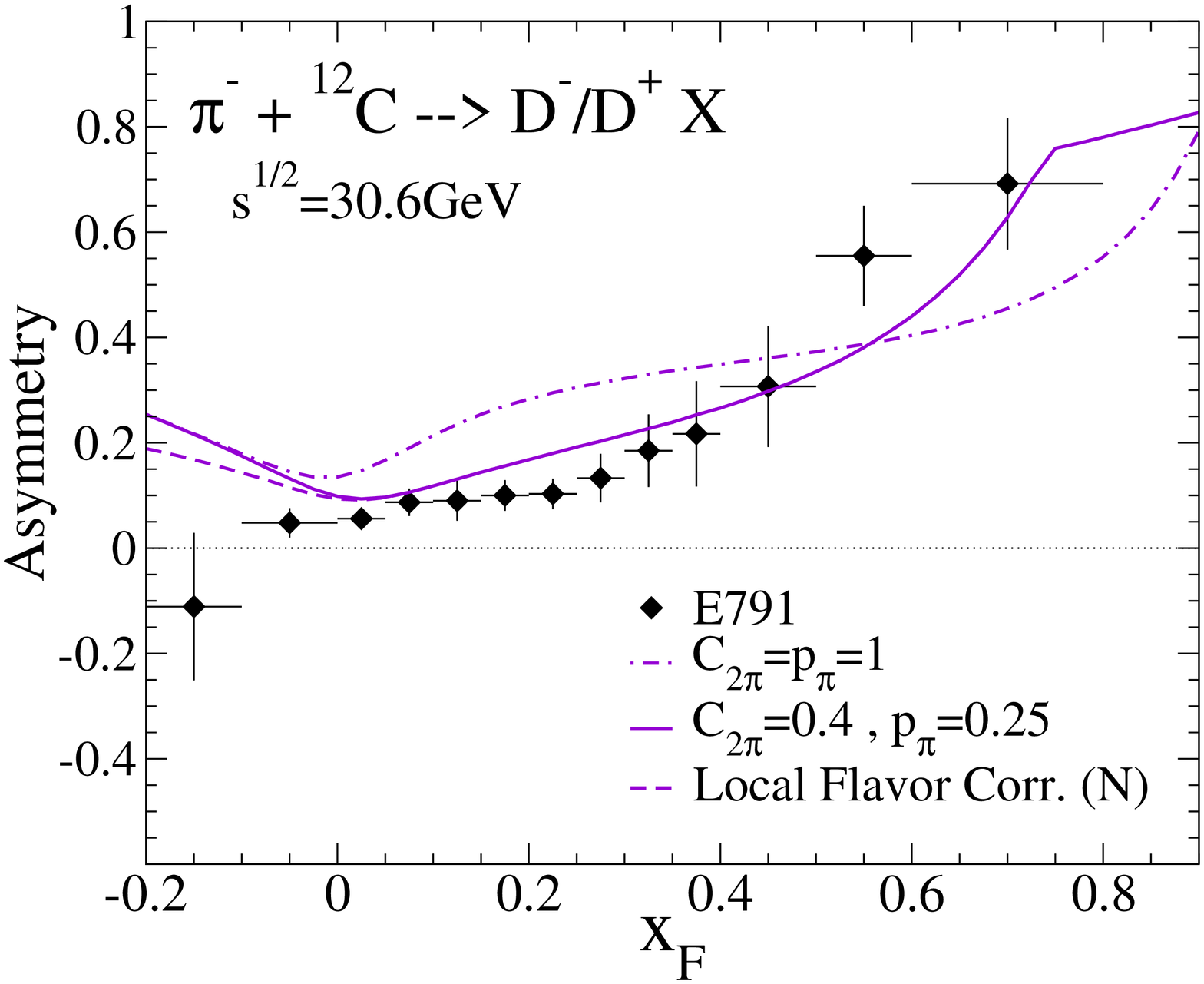,width=8cm,angle=-90}
\ece
\vspace{-0.2cm}
\caption{E791 data~\protect\cite{e791} for the $D^-/D^+$ asymmetry versus 
$x_F$ in 500~GeV $\pi^- A$ collisions, compared to recombination model 
calculations supplemented with $\delta$-function fragmentation contributions.}
\label{fig_Ae791}
\eef
At a higher projectile energy (500~GeV), the E791 collaboration
measured the $x_F$-dependence of the inclusive $D^-/D^+$ ratio.
As before, the modified 2-PDF with $C_2^\pi=0.4$ and $p_\pi=1/4$
accounts much better for the data than with $C_2^\pi=p_\pi=1$,
cf.~Fig.~\ref{fig_Ae791}.
In general, over the range of fixed target
energies discussed here, the projectile-energy variation in the
$x_F$-dependencies of the flavor asymmetries is rather weak.
\bef
\vspace{-0.8cm}
\bce
\epsfig{file=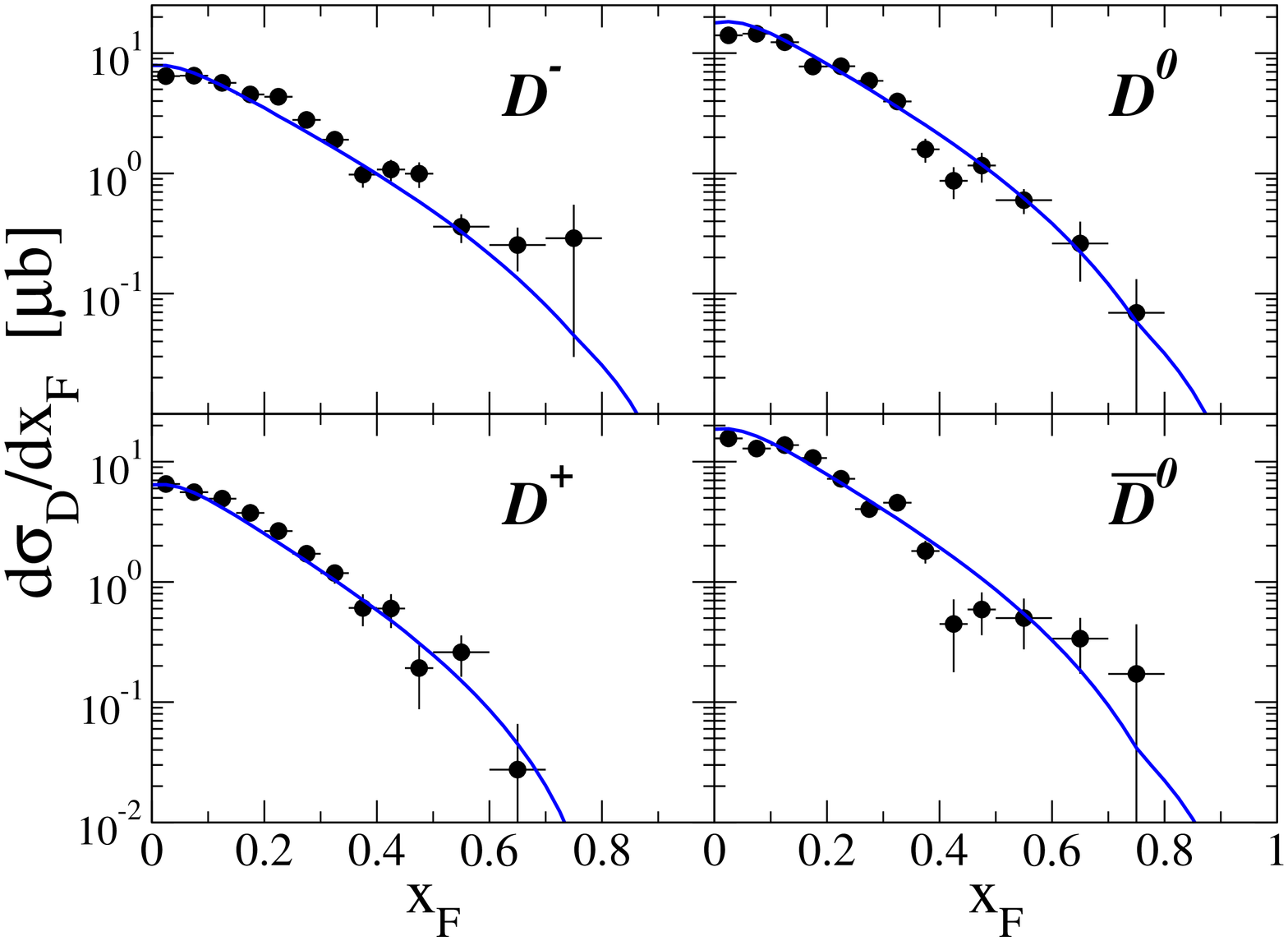,width=10cm,angle=-90}
\ece
\caption{$x_F$-differential cross sections from WA92~\protect\cite{wa92}
for inclusive yields of light $D$-mesons in 350~GeV $\pi^- A$ reactions, 
compared to recombination model calculations supplemented with fragmentation
contributions.}
\label{fig_Dwa-92}
\eef
In Fig.~\ref{fig_Dwa-92} differential $x_F$-spectra for the absolutely
normalized recombination + fragmentation model calculations (with $K$=3
in accordance with Fig.~\ref{fig_sigccb}) for the
individual $D$-meson states (underlying the solid lines in
Figs.~\ref{fig_Awa92} and \ref{fig_Ae791}) are compared to data 
from WA92~\cite{wa92}.  Reasonable agreement in all channels over the 
entire $x_F$-range is observed.

\subsubsection{$\sqrt{s}$ Dependencies}
We finally turn to the excitation function of the flavor asymmetries
in $\pi N$ collisions. Again, our combined fragmentation + recombination 
approach gives a rather satisfactory description of the observed 
ratios. In contrast to the $pN$ case (where, however, the data
accuracy is rather limited), now also the charged over neutral ratio
is in line with the data.  
Overall, the flavor asymmetries introduced by recombination processes
improve the agreement as compared to contributions from (isospin-symmetric) 
fragmentation alone (represented by the dotted lines) in all channels. 
\bef
\vspace{-0.5cm}
\bce
\epsfig{file=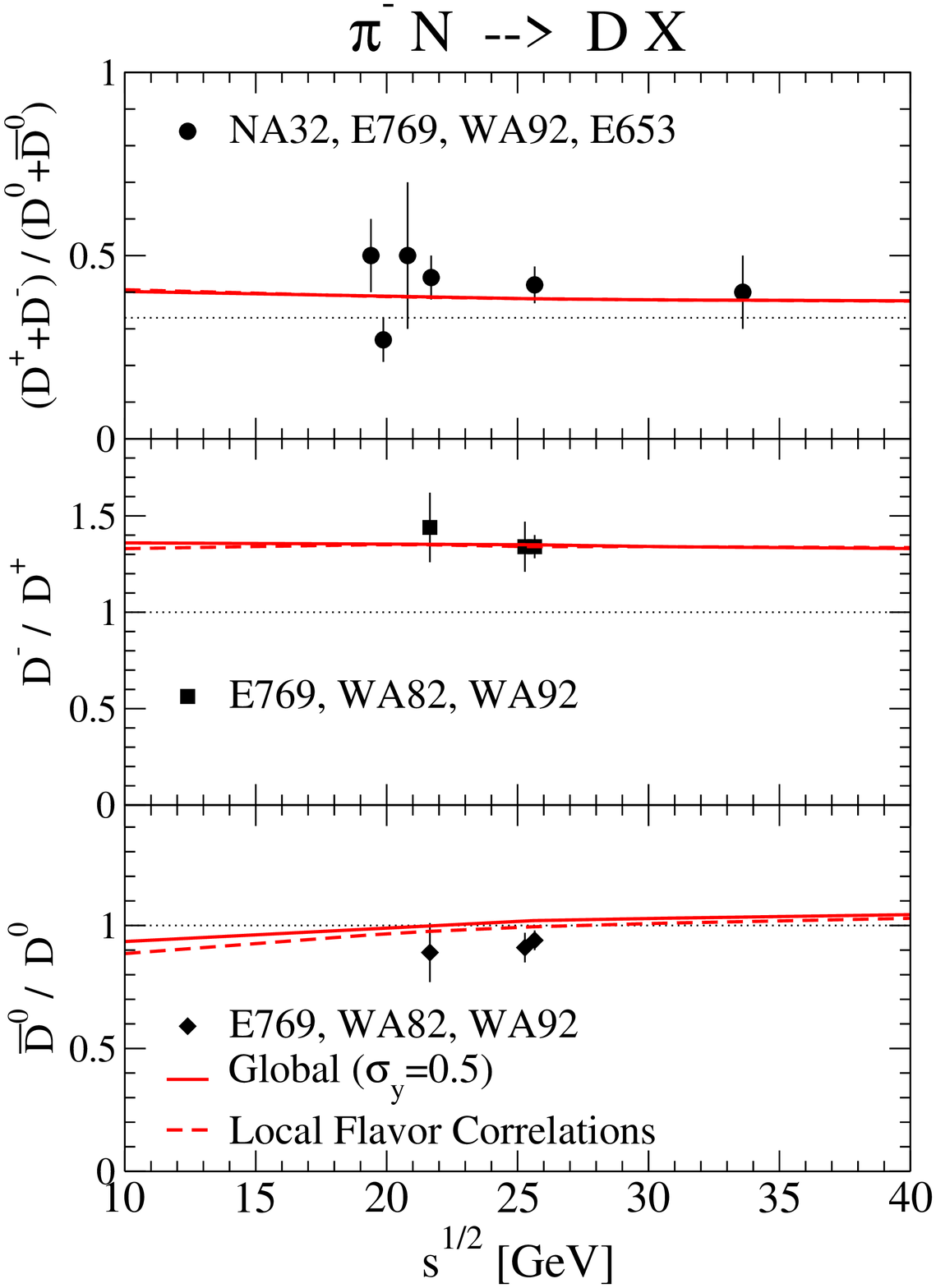,width=9cm,angle=0}
\ece
\caption{Collision energy dependence of various $D$-meson flavor
ratios in $\pi^- A$ collisions.}
\label{fig_ratpiN}
\eef

It should also be noted that overall charm conservation has not
been enforced explicitly. However, the slight excess of anti-charm
quarks induced by the $D^-/D^+$ asymmetry is conceivably  
compensated by a positive $\Lambda_c/\bar \Lambda_c$ asymmetry, 
which is also observed experimentally~\cite{selex-02} (but which we 
did not address here). The same argument applies for proton projectiles, 
where both $D^-/D^+$ and $\bar D^0/D^0$ are larger than one, but
at the same time the $\Lambda_c$ asymmetry is more pronounced than
for pion beams~\cite{selex-02}.   

We close this section with a note on $D_s$ mesons. For the 
recombination contribution alone, the ratio for (inclusive) strange
over nonstrange $D$-mesons at $\sqrt{s}=26$~GeV turns out to be
12\% for forward production (13\% for all $x_F$), with almost 
no energy dependence. This ratio increases slightly to 14-15\% 
when replacing the GRV-LO with the GRV-HO PDF's. 
From a compilation in Ref.~\cite{wa92}, the experimentally 
measured ratios can be found to be $15.8\pm2.9$\% for 
230~GeV $\pi^-$ beams~\cite{na32-91}, $16.7\pm3.3$\% at 
250~GeV~\cite{e769-96}, and $11.6\pm1.4$\%  at 350~GeV, 
with a combined  average of $R_{D_s/D}=12.9\pm 1.2$\%. 
This value is in surprising agreement with the recombination 
approach. In fact, for the remaining charm quarks, fragmentation 
should apply. The prediction of the LUND event generator~\cite{LUND}
(where hadronization is essentially performed by string fragmentation) 
turns out to be $\sim$8~\%, somewhat below the data. However, the 
combined recombination + fragmentation result would still be close
to the observed value.    

\section{Predictions for $p$-$p$ at RHIC}
\label{sec_rhic}
An important part of the physics program at the Relativistic 
Heavy-Ion Collider (RHIC) at Brookhaven consists of (polarized)
$p$-$p$ collisions up to a maximum {\it CM} of 500~GeV. 
Collisions at 200~GeV have been performed in the year-2 (2002) 
and year-3 running periods, although direct open charm
measurements are not expected before detector upgrades
have been performed. 
We here quote our predictions for $D$-meson
flavor ratios by extrapolating our recombination + fragmentation
approach according to Sect.~\ref{sec_pN} to 200~GeV {\it CM} 
energy~\footnote{We recall that the recombination part of the charm 
cross section increases with collision energy; at 200~GeV, the 
calculated fraction of $D$+$D_s$-mesons amounts to 67\% of the total 
$c\bar c$ yield (or 83\% of the expected total $D$+$D_s$ yield; note
that these numbers include the coefficient $C_2=0.8$ for the nucleon
2-PDF used in all results starting with Fig.~\ref{fig_ratpN}); our 
estimates for the ratios at RHIC energy 
do not account for potentially sizable destructive interference 
effects.}.  We find: \\ 
$R_{(D^++D^-)/(D^0+\bar D^0)}=0.40$ ; \  $R_{D^-/D^+}=1.24$ ; \
$R_{\bar D^0/ D^0}=1.35$ ; \ $R_{D_s/D}=0.23$ . \\  
These values are all rather close to the ones at fixed target 
energies. 

\section{Conclusions}
\label{sec_concl}
Charm-meson production in hadronic collisions at fixed target 
energies has been investigated with emphasis on asymmetries 
in the light-flavor sector. The calculations have been performed  
within a recombination model for produced anti-/charm quarks 
with surrounding light quarks in projectile and target, supplemented
with a standard fragmentation treatment for the remaining
$c$ and $\bar c$ quarks. More precisely, the recombination component
has been evaluated assuming a factorization of the 
perturbatively calculated (hard) $c\bar c$ production vertex 
and the subsequent (nonperturbative) "coalescence" process. 
On the one hand, this requires the use of a 2-parton distribution
function, for which we used a simple factorization ansatz (as first 
suggested 25 years ago). On the other hand, the recombination
function, as compared to previous analyses, has been generalized 
to incorporate coalescence with sea-quarks within the colliding 
hadrons, necessary to address 
bulk production dominated by the central region ($x_F\simeq 0$). 
Color correlations have been neglected throughout.
 
Employing recent parameterizations of parton distribution functions 
for nucleon target as well as pion and proton projectiles, we have 
found that (i) recombination processes contribute significantly to 
total $D$-meson yields and, (ii) the description of observed flavor 
asymmetries --  $(D^++D^-)/(D^0+\bar D^0)$, $D^-/D^+$ and 
$\bar D^0/D^0$ -- in bulk production is improved throughout as 
compared to isospin-symmetric fragmentation alone. The nontrivial 
isospin asymmetries (\ie, beyond known feeddown corrections from $D^*$
states) are driven by the valence- and sea-quark content of 
the colliding hadrons. No significant sensitivity was found with respect 
to {\it local} flavor correlations, as one might expect from 
valence-quark or quark-diquark substructures in the nucleon.      
Except for the charged-over-neutral asymmetry for proton projectiles
(which is, however, beset with large uncertainties), the magnitude
and  (not very pronounced) collision energy dependence of  
available fixed target data is rather well reproduced.  

As has been noted before, the same framework is also able to explain  
the differential $x_F$ dependencies of the flavor asymmetries as 
measured in $\pi N$ collisions. These become more pronounced at 
forward $x_F$, and are, within the model calculations, more sensitive 
to phase space correlations in the 2-parton distribution functions.  

Future data on identified charm mesons at collider energies, such as
expected from  $p$-$p$ runs at RHIC, will be most valuable to further 
test the relevance of recombination mechanisms in the hadronization
process, and thus contribute to a better understanding of the latter.

\vskip0.4cm
 
\centerline {\bf ACKNOWLEDGMENTS}
We gratefully acknowledge useful discussions with G. Sterman.  
This work was supported in part by the U.S. Department of Energy 
under Grant No. DE-FG02-88ER40388.

%
\end{document}